\documentclass[
superscriptaddress,
nofootinbib,
twocolumn,
 amsmath,amssymb,
 aps,
pra,
notitlepage,
longbibliography
]{revtex4-1}

\usepackage{dcolumn}
\usepackage{bm}
\usepackage{epsfig}
\usepackage{graphicx}
\usepackage{latexsym}
\usepackage{amsfonts}
\usepackage{setspace}
\usepackage{graphicx}
\usepackage{amsmath}
\usepackage{mathrsfs}
\usepackage{verbatim}
\usepackage{color}
\usepackage{SIunits}
\usepackage{hyperref}

\usepackage{physics}

\newcommand{\be}{\begin{equation}}
\newcommand{\ee}{\end{equation}}
\newcommand{\ba}{\begin{array}}
\newcommand{\ea}{\end{array}}
\newcommand{\bqa}{\begin{eqnarray}}
\newcommand{\eqa}{\end{eqnarray}}

\begin{document}

\title{Few-photon transport via a multimode nonlinear cavity: theory and applications}

\author{Yunkai Wang} 
\affiliation{Holonyak Micro and Nanotechnology Laboratory, University of Illinois at Urbana-Champaign, Urbana, IL 61801 USA}
\affiliation{Department of Physics, University of Illinois at Urbana-Champaign, Urbana, IL 61801 USA}
\affiliation{Illinois Quantum Information Science and Technology Center, University of Illinois at Urbana-Champaign, Urbana, IL 61801 USA}
\author{Kejie Fang} 
\email{kfang3@illinois.edu}
\affiliation{Holonyak Micro and Nanotechnology Laboratory, University of Illinois at Urbana-Champaign, Urbana, IL 61801 USA}
\affiliation{Illinois Quantum Information Science and Technology Center, University of Illinois at Urbana-Champaign, Urbana, IL 61801 USA}
\affiliation{Department of Electrical and Computer Engineering, University of Illinois at Urbana-Champaign, Urbana, IL 61801 USA}


\begin{abstract}

Few-photon transport via waveguide-coupled local quantum systems has attracted extensive theoretical and experimental studies. Most of the study has focused on atomic or atomic-like local quantum systems due to their strong light-matter interaction useful for quantum applications. Here, we study few-photon transport via a waveguide-coupled multimode optical cavity with second-order bulk nonlinearity. We develop a Feynman diagram approach and compute the scattering matrix of the one- and two-photon transport. Based on the calculated scattering matrix, we show highly nonclassical photonic effects, including photon blockade and $\pi-$conditional phase shift, are achievable in the waveguide-coupled multimode optical cavity system via quantum interference and linear response engineering. Our results might lead to significant applications of quantum photonic circuits in all-optical quantum information processing and quantum network protocols.

\end{abstract}

\maketitle

\section{Introduction}
The architecture of waveguide-coupled local quantum systems enables coupling flying photons with stationary qubits for light-matter interactions. A prime example is the waveguide quantum electrodynamics, with several types of local quantum systems demonstrated successfully, including trapped atoms \cite{tiecke2014nanophotonic,shomroni2014all,volz2014nonlinear}, solid state defects \cite{javadi2015single,sipahigil2016integrated}, and superconducting qubits \cite{mirhosseini2018superconducting,kannan2020waveguide}. This architecture is also relevant to quantum networks where optic fibers link local quantum nodes for routing, storage, and processing of quantum information encoded in individual photons \cite{simon2017towards}. In parallel, a large body of theoretical work has been carried out to study few-photon transport in the setting of waveguide-coupled local quantum systems \cite{chang2007single,shen2007strongly,liao2010correlated,zheng2010waveguide,rephaeli2012few,xu2015input}. Most of the theoretical and experimental work, though, has focused on qubit-like quantum entities, which are sought to provide strong light-matter interactions for quantum information tasks. 

Here, we study few-photon transport via a waveguide-coupled multimode optical cavity with second-order bulk nonlinearity. This is a common device setup used for cavity-enhanced parametric nonlinear optical processes including second-harmonic generation and parametric down-conversion \cite{lake2016efficient,guo2017parametric,luo2018highly,chang2019strong,zhang2019broadband,lukin20204h, lu2021efficient,zhao2021ingap}. However, its property at the few-photon level has not been fully explored, because bulk nonlinearity is generally believed to be insubstantial to impact single photons. We challenge this perspective by investigating the non-parametric few-photon transport and its physical observations. Besides using a systematic method from Ref. \cite{xu2015input} to compute the scattering matrix ($S-$matrix) of few-photon transport, we develop a Feynman diagram approach which provides further physical insight into the obtained $S-$matrix while offering mathematical simplicity in its computation. This method is particularly suited for weak nonlinear systems where only leading order Feynman diagrams are of interest. Exploiting the result of the $S-$matrix, we then study observational effects of few-photon transport via the waveguide-coupled multimode cavity, including photon blockade, conditional phase shift, and non-classical two-mode correlations. Surprisingly, while these effects are commonly believed to be associated with strong light-matter interaction, we find they could become substantial for waveguide-coupled optical cavities with only weak nonlinearity. 

These results reveal a general approach for creation and control of few-photon correlations via quantum interference and linear response engineering. The few-photon transport amplitude of waveguide-coupled local quantum systems is a superposition of the interaction-free, i.e., linear, and interaction-mediated amplitudes. When the linear transmission coefficient is comparable to the interaction-mediated amplitude, quantum interference between the two pathways leads to highly nonclassical correlations between the propagating photons. For weak nonlinear systems, this might need a tuned linear transmission which could be achieved, for example, in a one-port waveguide-cavity configuration. However, this approach is distinct from the post-selection method where photon-photon correlations are induced by the measurement.  Our result might lead to significant applications in quantum information science using quantum photonic circuits beyond the parametric regime, including quantum nondemolition measurement of photons \cite{xia2016cavity}, two-photon quantum logic gates \cite{langford2011efficient}, and nonlinearity-assisted entanglement swapping \cite{sangouard2011faithful}.

\section{Scattering matrix of few-photon transport}\label{Hamiltonian}

\begin{figure}[!tb]
\includegraphics[width=0.3\textwidth]{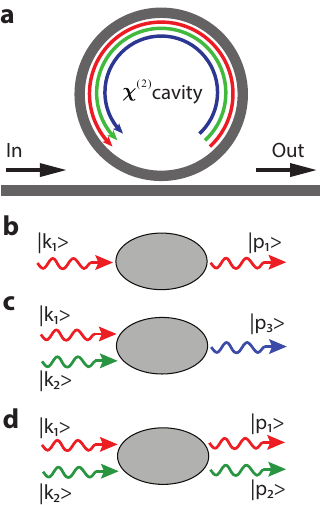}
\caption{\textbf{a}. A $\chi^{(2)}$ microring cavity with three resonances side-coupled to a waveguide. Red, green, and blue lines corresponds to $a_1$, $a_2$, and $a_3$ modes/photons, respectively. \textbf{b}-\textbf{d}. Illustrations of few-photon transport processes. $\ket{k}$ denotes the single-photon state with momentum $k$. }\label{cavity}
\end{figure}

The system under consideration is a waveguide-coupled trimodal optical cavity made from $\chi^{(2)}$ materials. A typical realization of such an optical cavity is a microring resonator (Fig. \ref{cavity}a) which supports traveling-wave resonances. These resonances couple via the $\chi^{(2)}$ nonlinearity when frequency-matching and, if necessary, phase-matching conditions are satisfied. Photons couple in and out of these resonances via the waveguide and meanwhile dissipate in other loss channels. In the Heisenberg-Langevin framework, this open quantum system can be modeled by the following effective Hamiltonian ($\hbar=1$)
\begin{widetext}
\begin{equation}\label{internal Hamiltonian}
H_{\textrm{eff}}=(\omega_1-i\frac{\kappa_{1}}{2})a^\dagger_1 a_1+(\omega_2-i\frac{\kappa_{2}}{2})a^\dagger_2 a_2+(\omega_3-i\frac{\kappa_{3}}{2})a^\dagger_3 a_3+g(a_1a_2a^\dagger_3+a^\dagger_1a^\dagger_2a_3),
\end{equation}
\end{widetext}
where $\omega_j$ and $\kappa_{j}$ are the frequency and photon loss rate of the $j$-th resonance, respectively, and $g$ is the trimodal coupling coefficient. For simplicity, we define a complex frequency $\alpha_j\equiv\omega_j-i\frac{\kappa_j}{2}$. For the trimodal interaction to be resonantly enhanced, the three modes need to satisfy the frequency-matching condition $\omega_1+\omega_2\approx \omega_3$. The cavity photon loss includes both leakage into the waveguide and intrinsic losses due to, for example, material absorption and surface scattering, i.e., $\kappa_{j}=\kappa_{je}+\kappa_{ji}$, where $\kappa_{je}$ is the cavity-waveguide coupling rate and $\kappa_{ji}$ is the intrinsic photon loss rate. Depending on the configuration of the waveguide-cavity coupling, the relation between the states of the outgoing and incoming photons in the waveguide is determined by the input-output formalism \cite{walls2007quantum}.  In this paper, we specifically consider the case of a cavity side-coupled to a waveguide unidirectionally---typical for ring resonators---which has the following input-output relation of the operators,
\begin{equation}\label{InOut}
a_{\textrm{out},j}(t)=a_{\textrm{in},j}(t)-i\sqrt{\kappa_{je}}a_{j}(t).
\end{equation} 
It turns out such one-port cavity-waveguide configuration is critical for realizing strong quantum correlations between propagating waveguide photons coupled via a weak nonlinear optical cavity. However, the computation method developed here can be straightforwardly generalized to other waveguide-cavity configurations. For the calculation below, we also assume a linear dispersion of the waveguide, which is justified given the linewidth of the cavity, and thus the operation bandwidth of the system, is much smaller than the cavity frequency \cite{fan2010input}.

The goal of this section is to calculate the scattering matrix ($S$-matrix) of the one- and two-photon transport via the waveguide-coupled trimodal $\chi^{(2)}$ cavity (Fig. \ref{cavity}b-d). We first use a systematic method from Ref. \cite{xu2015input} and apply it to the present case of a multi-mode $\chi^{(2)}$ cavity. We also develop a Feynman diagram approach which provides further physical insight of the photon transport while lessen the computation complexity associated with the non-perturbative method of Ref. \cite{xu2015input}, especially for weak nonlinear systems.

\subsection{One-to-one-photon transport}
We first consider the transport of a single photon (Fig. \ref{cavity}b). The amplitude of transport of an incoming photon at time $t'$ to an outgoing photon at time $t$ is given by the time-domain $S-$matrix
\be
S_{t;t'}\equiv\bra{0}a_{\textrm{out},1}(t)a_{\textrm{in},1}^\dagger(t')\ket{0},
\ee
where the photon is assumed to be in resonant with the $a_1$ mode.
As shown in Ref. \cite{xu2015input}, the time-domain $S-$matrix is related to the Green's function of the local system, i.e., here the nonlinear optical cavity, by applying Eq. \ref{InOut} and the quantum causality condition
\bqa
\label{QCC1}
\left[a(t), I\left(t^{\prime}\right)\right]=\left[a^{\dagger}(t), I\left(t^{\prime}\right)\right]=0, \quad \text { for } \quad t \leqslant t^{\prime}, \\
\left[a(t), O\left(t^{\prime}\right)\right]=\left[a^{\dagger}(t), O\left(t^{\prime}\right)\right]=0, \quad \text { for } \quad t \geqslant t^{\prime}, \label{QCC2}
\eqa
where $I(O)\left(t^{\prime}\right)$ is a shorthand notation for the input(output) operators that represent either $a_{\text {in(out) }}\left(t^{\prime}\right)$ or $a_{\text {in(out)}}^{\dagger}\left(t^{\prime}\right)$. Thus, the single-photon $S$-matrix becomes
\begin{equation}\label{S1t}
S_{t;t'}=\delta(t-t')-\kappa_{1e}G(t;t'),
\end{equation}
where $G(t;t')=\bra{0}\hat{T}[a_1(t)a_1^\dagger(t')]\ket{0}$ is the two-point Green's function, $\hat{T}$ is the time ordering operator and we have also used $[a_{\textrm{in(out)}}(t), a_{\textrm{in(out)}}^\dagger(t')]=\delta(t-t')$ in the derivation.

The Green's function of the local quantum system is calculated using operators time-evolved according to the effective Hamiltonian of \ref{internal Hamiltonian} \cite{xu2015input}. Thus, we proceed
\begin{equation}\label{Green1t}
\begin{aligned}
&G(t;t')\\
=&\bra{0}e^{iH_{\textrm{eff}}t}a_1e^{-iH_{\textrm{eff}}t}e^{iH_{\textrm{eff}}t'}a^\dagger_1e^{-iH_{\textrm{eff}}t'}\ket{0}\theta (t-t')\\
=&e^{i\alpha_1(t'-t)}\theta (t-t'),
\end{aligned}
\end{equation}
where $\theta(t-t')$ is the Heaviside step function. 

The frequency-domain $S$-matrix quantifying the transport amplitude between output and input photons with fixed frequency or momentum is given by the Fourier transform of the time-domain $S$-matrix. For the single-photon transport,
\begin{equation}\begin{aligned}\label{FFT}
S_{p_1;k_1}&\equiv \mathscr{F}\left\{  S_{t;t'} \right\} \\&= \int \frac{d t}{\sqrt{2 \pi}} e^{i p_{1} t} \int \frac{d t'}{\sqrt{2 \pi}} e^{i k_{1} t'} S_{t;t'} \\
& =\delta(p_1-k_1)-\kappa_{1e}G(p_1;k_1),
\end{aligned}\end{equation}
where $G(p_1;k_1)$ is the Fourier transform of the Green's function of \ref{Green1t}. This definition of Fourier transform leads to $\bra{k}\ket{k'}=\delta(k-k')$ given $\bra{t}\ket{t'}=\delta(t-t')$, where $\ket{k}\equiv \mathscr{F}\left\{ \ket{t}\right\}$. The momentum-space Green's function thus is
\begin{equation}\label{green_1}
G(p_1;k_1)=\frac{i}{k_1-\alpha_1}\delta(p_1-k_1).
\end{equation}
As a result, we have
\begin{equation}\label{scatter_1}
S_{p_1;k_1}=\left[ 1-\frac{i\kappa_{1e}}{k_1-\alpha_1}\right]\delta(p_1-k_1)\equiv t_{k_1}\delta(p_1-k_1),
\end{equation}
where $t_{k_1}$ is the single-photon transmission coefficient. The derivation above straightforwardly applies to transport of a single photon in resonant with other resonances. For the rest of the paper, we use the subscript $j$ in the momentum to indicate photons in resonant with the $j$-th resonance.

\subsection{Two-to-one-photon transport and vice versa}
Next, we consider the transport of two $a_1$ and $a_2$ photons to one $a_3$ photon enabled by the $\chi^{(2)}$ cavity (Fig. \ref{cavity}c). This corresponds to the sum-of-frequency process in nonlinear optics.  Following a similar procedure, the time-domain $S$-matrix of this process is found to be
\begin{widetext} 
\begin{equation}\begin{aligned}\label{S3tdef}
S_{t_3;t_1't_2'}&\equiv\bra{0}a_{\textrm{out},3}(t_3)a_{\textrm{in},1}^\dagger(t_1')a_{\textrm{in},2}^\dagger(t_2')\ket{0}\\
&=i\sqrt{\kappa_{1e}\kappa_{2e}\kappa_{3e}}\bra{0}\hat{T}[a_3(t_3)a_1^\dagger(t_1')a_2^\dagger(t_2')]\ket{0}\\
&= i\sqrt{\kappa_{1e}\kappa_{2e}\kappa_{3e}}\left(\bra{0}a_3(t_3)a^\dagger_1(t_1')a^\dagger_2(t_2')\ket{0}\theta(t_3-t_1')\theta(t_1'-t_2')+\bra{0}a_3(t_3)a^\dagger_2(t_2')a^\dagger_1(t_1')\ket{0}\theta(t_3-t_2')\theta(t_2'-t_1')\right).
\end{aligned}\end{equation}

We introduce $\ket{mnr}$ to denote the state with $m$, $n$, and $r$ photons in mode $a_1$, $a_2$, and $a_3$, respectively. One key property of the $\chi^{2}$ trimodal system is that $\{\ket{100}\}$, $\{\ket{010}\}$, and $\{\ket{110},\ket{001}\}$ form three closed subspaces of the effective Hamiltonian of \ref{internal Hamiltonian}. Based on this property, we proceed to compute Eq. \ref{S3tdef}:
\begin{equation}\label{S3t}
\begin{aligned}
&\bra{0}a_3(t_3)a^\dagger_1(t_1')a^\dagger_2(t_2')\ket{0}\\
=& \bra{000}a_3(t_3)\left(\ket{001}\bra{001}+\ket{110}\bra{110}\right) a_1^\dagger(t'_{1})\ket{010}\bra{010}a_2^\dagger(t'_{2})\ket{000}\\
=&\left(\bra{001}e^{-iH_{\textrm{eff}}t_3}\ket{001}\bra{001}e^{iH_{\textrm{eff}}t_1'}\ket{110}+\bra{001}e^{-iH_{\textrm{eff}}t_3}\ket{110}\bra{110}e^{iH_{\textrm{eff}}t_1'}\ket{110}\right)e^{-i\alpha_2(t_1'-t_2')}\\
=&\frac{g}{\sqrt{\Delta\alpha^2+4g^2}}e^{-i\alpha_2(t_1'-t_2')}(e^{i\lambda_2 (t_1'-t_3)}-e^{i\lambda_1(t_1'-t_3)}),
\end{aligned}
\end{equation}
where $\Delta\alpha=\alpha_1+\alpha_2-\alpha_3$ and $\lambda_{1,2}=\frac{1}{2}(\alpha_1+\alpha_2+\alpha_3)\pm\frac{1}{2}\sqrt{\Delta\alpha^2+4g^2}$ are the eigenvalues of $H_{\textrm{eff}}$ in the subspace spanned by $\ket{110}$ and $\ket{001}$, i.e., 
\begin{equation}
\left[ \begin{array}{cc}
\alpha_1+\alpha_2 & g\\
g & \alpha_3
\end{array} 
\right ].
\end{equation}
The result of $\bra{0}a_3(t_3)a^\dagger_2(t_2')a^\dagger_1(t_1')\ket{0}$ is obtained by exchanging the subscripts in Eq. \ref{S3t}. Finally, the momentum-space $S-$matrix, $S_{p_3;k_1k_2}\equiv \mathscr{F}\left\{ S_{t_3;t_1't_2'}\right\}$, is given by
\begin{equation}\label{Sp3}
S_{p_3;k_1k_2}\equiv i\sqrt{\kappa_{1e}\kappa_{2e}\kappa_{3e}}G(p_3;k_1,k_2)=-ig\frac{\sqrt{\kappa_{1e}\kappa_{2e}\kappa_{3e}}}{\sqrt{2\pi}}\frac{(k_1+k_2-\alpha_1-\alpha_2)}{(k_1-\alpha_1)(k_2-\alpha_2)(k_1+k_2-\lambda_1)(k_1+k_2-\lambda_2)}\delta(p_3-k_1-k_2),
\end{equation}
where $G(p_3;k_1,k_2)$ is the momentum-space Green's function.

For the degenerate case, i.e., $a_1$ and $a_2$ modes are the same mode, the matrix form of $H_{\textrm{eff}}$ in the subspace spanned by $\ket{110}$ (i.e., $\ket{20}$) and $\ket{001}$ is given by
\begin{equation}\label{heffd}
\left[ \begin{array}{cc}
2\alpha_1 & \sqrt{2}g\\
\sqrt{2}g & \alpha_3
\end{array} 
\right ].
\end{equation}
After a similar derivation, we obtain 
\begin{equation}\label{Sp3d}
S_{p_3;k_1k_2}=-2ig\frac{\sqrt{\kappa_{1e}^2\kappa_{3e}}}{\sqrt{2\pi}}\frac{(k_1+k_2-2\alpha_1)}{(k_1-\alpha_1)(k_2-\alpha_1)(k_1+k_2-\lambda'_1)(k_1+k_2-\lambda'_2)}\delta(p_3-k_1-k_2),
\end{equation}
\end{widetext}
where $\lambda'_{1,2}=\frac{1}{2}(2\alpha_1+\alpha_3)\pm\frac{1}{2}\sqrt{(2\alpha_1-\alpha_3)^2+8g^2}$ are the eigenvalues of the matrix of \ref{heffd}. 
It is worth pointing out that $\lambda_{1,2}$ ($\lambda'_{1,2}$) are functions of the trimodal coupling coefficient $g$, and thus Eqs. \ref{Sp3} and \ref{Sp3d} are non-perturbation result in terms of $g$, contributed by all virtual processes. These processes involving creation and annihilation of virtual photons will be revealed using the Feynman diagram approach in Section \ref{Sec:Feynman}.   

The reverse process of one-to-two-photon transport has an $S-$matrix
\be
S_{t_1't_2';t_3}\equiv\bra{0}a_{\textrm{out},1}(t_1')a_{\textrm{out},2}(t_2')a^\dagger_{\textrm{in},3}(t_3)\ket{0},
\ee
whose Fourier transform satisfies
\be\label{S1to2}
S_{k_1k_2;p_3}\equiv \mathscr{F}\left\{ S_{t_1't_2';t_3} \right\}=S_{p_3;k_1k_2}.
\ee

\subsection{Two-to-two-photon transport}
Last, we consider the process of transport of two $a_1$ and $a_2$ photons to two $a_1$ and $a_2$ photons, without conversion to $a_3$ photons (Fig. \ref{cavity}d). We again start from the time-domain $S$-matrix
\begin{widetext}
\begin{equation}\label{S4t}
\begin{aligned}
S_{t_1t_2;t'_1t'_2}&\equiv\bra{0}a_{\textrm{out},1}(t_1)a_{\textrm{out},2}(t_2)a_{\textrm{in},1}^\dagger(t_1')a_{\textrm{in},2}^\dagger(t_2')\ket{0}\\
&=\delta(t_1-t_1')\delta(t_2-t_2')-\kappa_{1e}\bra{0}\hat{T}[a_1(t_1)a_1^\dagger(t_1')]\ket{0}\delta(t_2-t_2')-\kappa_{2e}\bra{0}\hat{T}[a_2(t_2)a_2^\dagger(t_2')]\ket{0}\delta(t_1-t_1')\\
&\quad+\kappa_{1e}\kappa_{2e}\bra{0}\hat{T}[a_1(t_1)a_1(t_2)a_1^\dagger(t_1')a_2^\dagger(t_2')]\ket{0}.
\end{aligned}
\end{equation}
The second and third terms are the single-photon Green's function which have been calculated (see Eq. \ref{Green1t}). The last term is new and involves six different time orderings,
\begin{equation}\label{six terms}
\begin{aligned}
\bra{0}\hat{T}\left[a_2(t_2)a_1(t_1)a_{1}^\dagger(t_1')a_{2}^\dagger(t_2')\right]\ket{0}
=&\bra{0}a_2(t_{2})a_1(t_{1})a_1^\dagger(t'_{1})a_2^\dagger(t'_{2})\ket{0}\theta(t_2-t_1)\theta(t_1-t_1')\theta(t_1'-t_2)\\
&+\bra{0}a_2(t_2)a_1(t_1)a_2^\dagger(t_2') a_1^\dagger(t_1')\ket{0}\theta(t_2-t_1)\theta(t_1-t_2')\theta(t_2'-t_1')\\
&+\bra{0}a_1(t_1)a_1^\dagger(t_1') a_2(t_2) a_2^\dagger(t_2')\ket{0}\theta(t_1-t_1')\theta(t_1'-t_2)\theta(t_2-t_2')\\
&+\bra{0}a_1(t_1)a_2(t_2)a_2^\dagger(t_2') a_1^\dagger(t_1')\ket{0}\theta(t_1-t_2)\theta(t_2-t_2')\theta(t_2'-t_1)\\
&+\bra{0}a_1(t_{1})a_2(t_{2})a_1^\dagger(t'_{1})a_2^\dagger(t'_{2})\ket{0}\theta(t_1-t_2)\theta(t_2-t_1')\theta(t_1'-t_2')\\
&+\bra{0}a_2(t_2)a_2^\dagger(t_2') a_1(t_1) a_1^\dagger(t_1')\ket{0}\theta(t_2-t_2')\theta(t_2'-t_1)\theta(t_1-t_1').
\end{aligned}
\end{equation}
We first calculate $\bra{0}a_2(t_{2})a_1(t_{1})a_1^\dagger(t'_{1})a_2^\dagger(t'_{2})\ket{0}$:
\begin{equation}\label{6terms1}
\begin{aligned}
&\bra{0}a_2(t_{2})a_1(t_{1})a_1^\dagger(t'_{1})a_2^\dagger(t'_{2})\ket{0}\\
=&\bra{000}a_2(t_{2})\ket{010}\bra{010}a_1(t_{1})\left(\ket{110}\bra{110}+\ket{001}\bra{001}\right) a_1^\dagger(t'_{1})\ket{010}\bra{010}a_2^\dagger(t'_{2})\ket{000}\\
=&e^{i\alpha_2(-t_{2}+t'_{2}+t_{1}-t'_{1})}\big(\bra{110}e^{-iH_{\textrm{eff}}t_{1}}\ket{110}\bra{110}e^{iH_{\textrm{eff}}t'_{1}}\ket{110}+\bra{110}e^{-iH_{\textrm{eff}}t_{1}}\ket{001}\bra{001}e^{iH_{\textrm{eff}}t'_{1}}\ket{110}\big)\\
=&e^{i\alpha_2(-t_2+t'_2+t_1-t'_1)}\bigg[\Big(\frac{1}{2}-\frac{1}{2}\frac{\Delta\alpha}{\sqrt{\Delta\alpha^2+4g^2}}\Big)e^{-i\lambda_1(t_1-t'_1)}+\Big(\frac{1}{2}+\frac{1}{2}\frac{\Delta\alpha}{\sqrt{\Delta\alpha^2+4g^2}}\Big)e^{-i\lambda_2(t_1-t'_1)}\bigg].
\end{aligned}
\end{equation}
Similarly, the second and third terms in Eq. \ref{six terms} are found to be
\begin{equation}\label{6terms2}
\bra{0}a_2(t_2)a_1(t_1)a_2^\dagger(t_2') a_1^\dagger(t_1')\ket{0}=e^{i\alpha_2(t_1-t_2)}e^{i\alpha_1(t'_1-t'_2)}
\bigg[\Big(\frac{1}{2}-\frac{1}{2}\frac{\Delta\alpha}{\sqrt{\Delta\alpha^2+4g^2}}\Big)e^{-i\lambda_1(t_1-t'_2)}+\Big(\frac{1}{2}+\frac{1}{2}\frac{\Delta\alpha}{\sqrt{\Delta\alpha^2+4g^2}}\Big)e^{-i\lambda_2(t_1-t'_2)}\bigg],
\end{equation}
\begin{equation}\label{6terms3}
\bra{0}a_1(t_1)a_1^\dagger(t_1') a_2(t_2) a_2^\dagger(t_2')\ket{0}= e^{i\alpha_1(t'_1-t_1)}e^{i\alpha_2(t'_2-t_2)}.
\end{equation}
The rest three terms in Eq. \ref{six terms} are obtained by simply exchanging the indices $1$ and $2$ in Eqs. \ref{6terms1}-\ref{6terms3}. 

Finally, by Fourier transform of Eq. \ref{six terms}, we obtain the momentum-space Green's function
\begin{equation}\label{G22}
G(p_1,p_2;k_1,k_2)= -\frac{1}{(k_1-\alpha_1)(k_2-\alpha_2)}\delta(p_1-k_1)\delta(p_2-k_2)+M(p_1,p_2,k_1,k_2)\delta(p_1+p_2-k_1-k_2),
\end{equation}
where
\begin{equation}\label{Tdnd}
M(p_1,p_2,k_1,k_2)=-\frac{ig^2}{2\pi}\frac{k_1+k_2-\alpha_1-\alpha_2}{(k_1-\alpha_1)(k_2-\alpha_2)(p_1-\alpha_1)(p_2-\alpha_2)(k_1+k_2-\lambda_1)(k_1+k_2-\lambda_2)},
\end{equation}
and the momentum-space $S-$matrix
\begin{equation}\label{S22}
S_{p_1p_2;k_1k_2}= t_{k_1}t_{k_2}\delta(p_1-k_1)\delta(p_2-k_2)+\kappa_{1e}\kappa_{2e}M(p_1,p_2,k_1,k_2)\delta(p_1+p_2-k_1-k_2).
\end{equation}

For the case of degenerate $a_1$ and $a_2$, after a similar derivation, we have
\begin{equation}\label{G22d}
G(p_1,p_2;k_1,k_2)= -\frac{1}{(k_1-\alpha_1)(k_2-\alpha_1)}[\delta(p_1-k_1)\delta(p_2-k_2)+\delta(p_1-k_2)\delta(p_2-k_1)]+M(p_1,p_2,k_1,k_2)\delta(p_1+p_2-k_1-k_2)
\end{equation}
and
\begin{equation}\label{S22d}
S_{p_1p_2;k_1k_2}= t_{k_1}t_{k_2}[\delta(p_1-k_1)\delta(p_2-k_2)+\delta(p_1-k_2)\delta(p_2-k_1)]+\kappa_{1e}^2M(p_1,p_2,k_1,k_2)\delta(p_1+p_2-k_1-k_2),
\end{equation}
where
\begin{equation}\label{Td}
M(p_1,p_2,k_1,k_2)=-\frac{2ig^2}{\pi}\frac{k_1+k_2-2\alpha_1}{(k_1-\alpha_1)(k_2-\alpha_1)(p_1-\alpha_1)(p_2-\alpha_1)(k_1+k_2-\lambda'_1)(k_1+k_2-\lambda'_2)}.
\end{equation}
\end{widetext}
Eqs. \ref{S22} and \ref{S22d} show that the $S-$matrix of two-to-two-photon transport consists of an interaction-free component, which is the product of two single-photon transmission coefficients, and an interaction-mediated component. The total transport amplitude is given by the superposition of the two components, representing the quantum interference between the two pathways. This leads to profound consequences of the photon-photon interaction via the waveguide-coupled nonlinear optical cavities and applications in quantum information science (see Section \ref{result}). The interaction-mediated component was also interpreted as the two-photon bound state previously \cite{shen2007strongly}.

\section{Feynman diagram approach}\label{Sec:Feynman}

\begin{figure*}[!tb]
\includegraphics[width=\textwidth]{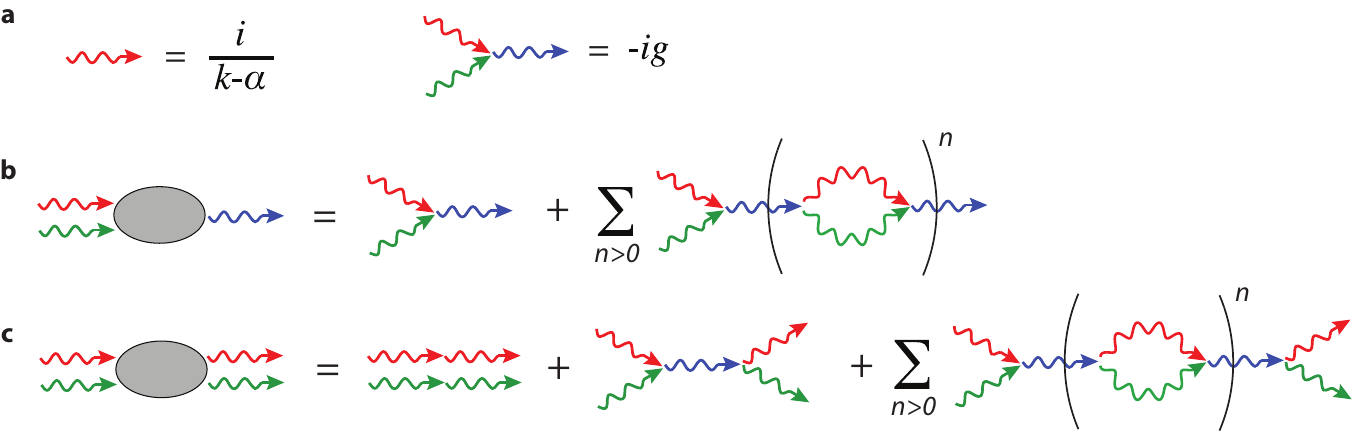}
\caption{ \textbf{a}. Feynman diagram rules for the propagator and vertex. Red, green, and blue lines corresponds to $a_1$, $a_2$, and $a_3$ mode photons, respectively. \textbf{b} and \textbf{c}. Feynman diagrams contributing to the Green's function of two-to-one-photon transport (\textbf{b}) and two-to-two-photon transport (\textbf{c}). The part enclosed by the parenthesis is repeated by $n$ times. }\label{Feynman}
\end{figure*}

In this section, we introduce a perturbation method based on Feynman diagrams to calculate the momentum-space Green's function and $S-$matrix. The perturbation method, suited for weak nonlinear systems with $g/\kappa_j<1$, reveals the relevant physical processes contributing to the few-photon transport amplitude.  It is also mathematically simplified, especially when only the leading order Feynman diagrams are of interest, compared to the previous method where careful cancellation of divergence is needed when performing the Fourier transform of the time-domain Green's function involving the Heaviside step function.  The Feynman diagram rules here resemble those developed in quantum filed theory (see, e.g., Ref. \cite{peskin2018introduction}). In essence, the $n-$operator Green's function is calculated using all connected Feynman diagrams with $n$ external points, which are constructed from basic elements including ``propagator'' and ``vertex''. The propagator is the two-point free-particle Green's function and the vertex corresponds to the bare interaction term of the Hamiltonian. For the Hamiltonian of \ref{internal Hamiltonian}, the Feynman diagram rules are given as follows:
\begin{itemize}
\item Propagator: $\frac{i}{k-\alpha}$,
\item Vertex: $-ig$,
\item Impose energy conservation at each vertex: $\frac{1}{\sqrt{2\pi}}\delta(\sum k-\sum p)$,
\item Integrate undetermined momentum: $\int dk\int dp$. 
\item Multiply the symmetry factor: $m!$ for $m$ propagators of the same mode connecting two vertices or connecting a vertex with external points. 
\end{itemize}

The Feynman diagrams corresponding to the propagator and vertex are shown in Fig. \ref{Feynman}a. A few differences from the Feynman diagram rules for field operators in quantum filed theory are worth noting. Here, because the equation of motion for the cavity mode operator is a linear differential equation, the propagator, given by Eq. \ref{green_1}, is linearly dependent on the momentum. Further, because the effective Hamiltonian of the open quantum system already includes loss terms, i.e., $i\kappa_j$, the propagator does not diverge at the real resonance frequency, avoiding the addition of an infinitesimal imaginary term in the propagator for performing the momentum integrals.

The three- and four-point momentum-space Green's functions related to Eqs. \ref{S3tdef} and \ref{six terms} can be calculated using the Feynman diagrams shown in Fig. \ref{Feynman}b and c, respectively. These Feynman diagrams also illustrates the physical processes, involving creation and annihilation of virtual photons, that contribute to the photon transport amplitude. After computation of each Feynman diagram according to the rules above, the three-point Green's function is found to be
\begin{widetext}
\be\label{FG3}
G(p_3;k_1,k_2)=-\frac{g}{\sqrt{2\pi}}\frac{1}{(k_1-\alpha_1)(k_2-\alpha_2)(p_3-\alpha_3)}\sum_{n=0}^{\infty} \left( \frac{g^2}{(p_3-\alpha_1-\alpha_2)(p_3-\alpha_3)}  \right)^n\delta(p_3-k_1-k_2),
\ee 
\end{widetext}
where the term of $n=0$ corresponds to the first diagram on the right hand side in Fig. \ref{Feynman}b and each term of $n>0$ corresponds to the diagram with $n$ repeated part in the parenthesis.
The four-point Green's function is found to be
\begin{widetext}
\begin{equation}
\begin{aligned}\label{FG22}
&G(p_1,p_2;k_1,k_2)\\=&-\frac{1}{(k_1-\alpha_1)(k_2-\alpha_2)}\delta(p_1-k_1)\delta(p_2-k_2)\\&-\frac{ig^2}{2\pi}\frac{1}{(k_1-\alpha_1)(k_2-\alpha_2)(p_1-\alpha_1)(p_2-\alpha_2)(k_1+k_2-\alpha_3)}\sum_{n=0}^{\infty} \left( \frac{g^2}{(k_1+k_2-\alpha_1-\alpha_2)(k_1+k_2-\alpha_3)} \right)^n \delta(p_1+p_2-k_1-k_2),
\end{aligned}
\end{equation}
\end{widetext}
where the first term corresponds to the first diagram on the right hand side in Fig. \ref{Feynman}c, the term of $n=0$ corresponds to the second diagram, and each term of $n>0$ corresponds to the diagram with $n$ repeated part in the parenthesis.
One can confirm that Eqs. \ref{FG3} and \ref{FG22} reproduce Eqs. \ref{Sp3} and \ref{G22}, respectively, after performing the summation. As an example of applying the Feynman diagram rules, we explicitly show the calculation of the $n=1$ diagram in Fig. \ref{Feynman}b. Higher order diagrams are essentially multiple repetition of the component in the parenthesis of this basic diagram. The formula corresponding to this diagram, following the Feynman diagram rules from left to right, is given by
\begin{widetext}
\begin{equation}
\begin{aligned}
&\frac{i}{k_1-\alpha_1}\frac{i}{k_2-\alpha_2}\frac{-ig}{\sqrt{2\pi}}\int dq_3\delta(q_3-k_1-k_2)\frac{i}{q_3-\alpha_3}\frac{-ig}{\sqrt{2\pi}}\int dq_1dq_2\delta(q_1+q_2-q_3)\frac{i}{q_1-\alpha_1}\frac{i}{q_2-\alpha_2}\frac{-ig}{\sqrt{2\pi}}\delta(q_1+q_2-p_3)\frac{i}{p_3-\alpha_3}\\
=&-\frac{g}{\sqrt{2\pi}}\delta(p_3-k_1-k_2)\frac{1}{(k_1-\alpha_1)(k_2-\alpha_2)(p_3-\alpha_3)}\frac{-g^2}{2\pi i}\int dq_1dq_2\delta(q_1+q_2-p_3)\frac{1}{q_1-\alpha_1}\frac{1}{q_2-\alpha_2} \frac{1}{p_3-\alpha_3}\\
=& -\frac{g}{\sqrt{2\pi}}\delta(p_3-k_1-k_2)\frac{1}{(k_1-\alpha_1)(k_2-\alpha_2)(p_3-\alpha_3)}\frac{g^2}{(p_3-\alpha_1-\alpha_2)(p_3-\alpha_3)},
\end{aligned}
\end{equation}
\end{widetext}
which yields the $n=1$ term in Eq. \ref{FG3}. 

We note Eqs. \ref{FG3} and \ref{FG22} are power series expansion in terms of $(g/\kappa)^2$, where $\kappa$ is the shorthanded notation for $\kappa_j$ or their sum, for input and output photons in resonant with the cavity modes, i.e., $k_j, p_j\approx \omega_j$, and a frequency-matched cavity ($\omega_1+\omega_2\approx\omega_3$). As a result, only leading order Feynman diagrams are needed for weak nonlinear optical cavities, i.e., $(g/\kappa)^2< 1$, which significantly simplifies the calculation. The computation here is also divergence free, in contrast to the method used in the previous Section which encounters divergences during Fourier transform of the time-domain Green's function and thus requires careful cancellations \cite{xu2015input} 

\section{Observational effects of few-photon transport}\label{result}
In this Section, we discuss several observational effects associated with the few-photon transport via a waveguide-coupled $\chi^{(2)}$ cavity. Throughout this Section, we consider input photons in a weak coherent state, which are commonly used for single-photon level  experiments. 

\subsection{Single-photon down-conversion}
We first consider the down-conversion process of $a_3$ photons to $a_1$ and $a_2$ photons. The input weak coherent state of monochromatic $a_3$ photons can be written as
\begin{equation}\label{weakalpha}
\ket{\psi_{\textrm{in}}}=\ket{0}+\alpha\ket{1_t}+\frac{\alpha^2}{\sqrt{2}}\ket{1_t1_t}+O(\alpha^3),
\end{equation}
where $\alpha$ is the amplitude of the coherent state and $\ket{1_t}=\ket{1}e^{-ikt}$ represents a monochromatic single-photon state with momentum $k$ in the time domain. Note $\ket{1_t}\neq \ket{t}$. We are interested in the second-order correlation function of the output  $a_1$ and $a_2$ photons:
\begin{equation}
\begin{aligned}\label{downg2}
&g^{(2)}(\tau)\\=&\frac{\bra{\psi_{\textrm{in}}}a_{\textrm{out},1}^\dagger(t)a_{\textrm{out},2}^\dagger(t+\tau)a_{\textrm{out},2}(t+\tau)a_{\textrm{out},1}(t)\ket{\psi_{\textrm{in}}}}{\bra{\psi_{\textrm{in}}}a_{\textrm{out},1}^\dagger(t)a_{\textrm{out},1}(t)\ket{\psi_{\textrm{in}}}\bra{\psi_{\textrm{in}}}a_{\textrm{out},2}^\dagger(t)a_{\textrm{out},2}(t)\ket{\psi_{\textrm{in}}}},
\end{aligned}
\end{equation}
which can be calculated using the momentum-space $S$-matrix. To do so, we first perform a Fourier transform of $\ket{\psi_{\textrm{in}}}$:
\be
\mathscr{F}\left\{ \ket{\psi_{\textrm{in}}} \right\}=\int \frac{d t}{\sqrt{2 \pi}} e^{i k' t}  \ket{1}e^{-ikt}=\sqrt{2\pi}\ket{1}\delta(k'-k).
\ee
Thus we define Fourier-transformed monochromatic single-photon state to be $\sqrt{2\pi}\ket{1_{k}}$. 
Keeping the leading order term in $\ket{\psi_{\textrm{in}}}$, we have\begin{widetext}
\begin{equation}\begin{aligned}
&\bra{\psi_{\textrm{in}}}a_{\textrm{out},1}^\dagger(t)a_{\textrm{out},2}^\dagger(t+\tau)a_{\textrm{out},2}(t+\tau)a_{\textrm{out},1}(t)\ket{\psi_{\textrm{in}}}\\
=&|\alpha|^2\bra{1_t}a_{\textrm{out},1}^\dagger(t)a_{\textrm{out},2}^\dagger(t+\tau)a_{\textrm{out},2}(t+\tau)a_{\textrm{out},1}(t)\ket{1_{t}}\\
=&|\alpha|^2\bra{1_t}a_{\textrm{out},1}^\dagger(t)a_{\textrm{out},2}^\dagger(t+\tau)\ket{0}\bra{0}a_{\textrm{out},2}(t+\tau)a_{\textrm{out},1}(t)\ket{1_{t}}\\
=&\frac{|\alpha|^2}{2\pi}\int dp_1dp_2 \bra{1_{k}}a_{\textrm{out},1}^\dagger(p_1)a_{\textrm{out},2}^\dagger(p_2)\ket{0}  e^{ip_1t+ip_2(t+\tau)}\\
&\quad\times\int d\tilde p_1d\tilde p_2\bra{0}a_{\textrm{out},2}(\tilde p_2)a_{\textrm{out},1}(\tilde p_1)\ket{1_{k}}e^{-i\tilde p_1t-i\tilde p_2(t+\tau)}\\
=& \frac{|\alpha|^2}{2\pi}\int dp_1dp_2 S_{p_1p_2;k}^* e^{ip_1t+ip_2(t+\tau)}\\
&\quad \times \int d\tilde p_1d\tilde p_2 S_{\tilde p_1\tilde p_2;k}e^{-i\tilde p_1t-i\tilde p_2(t+\tau)} \\
=&\begin{cases}
\frac{|\alpha|^2g^2\kappa_{1e}\kappa_{2e}\kappa_{3e}}{|(k-\lambda_1)(k-\lambda_2)|^2}e^{-\kappa_2\tau},\tau>0, \\
\frac{|\alpha|^2g^2\kappa_{1e}\kappa_{2e}\kappa_{3e}}{|(k-\lambda_1)(k-\lambda_2)|^2}e^{\kappa_1\tau},\tau<0,
\end{cases}
\end{aligned}\end{equation}
where $S_{p_1p_2;k}$ is given by Eq. \ref{S1to2}, and 
\begin{equation}\begin{aligned}\label{flux}
&\bra{\psi_{\textrm{in}}}a_{\textrm{out},1}^\dagger(t)a_{\textrm{out},1}(t)\ket{\psi_{\textrm{in}}}\\
=&|\alpha|^2\bra{1_t}a_{\textrm{out},1}^\dagger(t)a_{\textrm{out},1}(t)\ket{1_{t}}\\
=&|\alpha|^2\int dp_1d\tilde p_1 \bra{1_{k}}a_{\textrm{out},1}^\dagger(p_1)a_{\textrm{out},1}(\tilde p_1)\ket{1_{k}}  e^{ip_1t-i\tilde p_1t}\\
=&|\alpha|^2\int dp_1d\tilde p_1\bra{1_{k}}a_{\textrm{out},1}^\dagger(p_1)\left(\int dp_2\ket{p_2}\bra{p_2}+\int dp_{2i}\ket{p_{2i}}\bra{p_{2i}}\right)a_{\textrm{out},1}(\tilde p_1)\ket{1_{k}}  e^{ip_1t-i\tilde p_1t}\\
=& |\alpha|^2\left(\int dp_1d\tilde p_1dp_2 S_{p_1p_2;k}S_{\tilde p_1p_2;k}^* +\frac{\kappa_{2i}}{\kappa_{2e}}\int dp_1d\tilde p_1dp_{2i} S_{p_1p_{2i};k}S_{\tilde p_1p_{2i};k}^* \right)e^{ip_1t-i\tilde p_1t}\\
=& |\alpha|^2\frac{\kappa_{2}}{\kappa_{2e}}\int dp_1d\tilde p_1dp_2 S_{p_1p_2;k}S_{\tilde p_1p_2;k}^*e^{ip_1t-i\tilde p_1t}\\
=&|\alpha|^2\frac{\kappa_{1e}}{\kappa_1}\frac{g^2(\kappa_1+\kappa_2)\kappa_{3e}}{|(k-\lambda_1)(k-\lambda_2)|^2},
\end{aligned}\end{equation}
\end{widetext}
where $\ket{p_{2i}}$ represents the state of $a_2$ photons in the intrinsic loss channel. Because the states in the waveguide and the intrinsic loss channel combined form a complete Hilbert space of the leaked photons from the cavity, we have the identity 
\be\label{identity}
\int dp_2\ket{p_2}\bra{p_2}+\int dp_{2i}\ket{p_{2i}}\bra{p_{2i}}=I.
\ee
In other words, $\bra{\psi_{\textrm{in}}}a_{\textrm{out},1}^\dagger(t)a_{\textrm{out},1}(t)\ket{\psi_{\textrm{in}}}$ measures the total $a_1$ photon flux, regardless of the place of $a_2$ photons, which could either be in the waveguide or the intrinsic loss channel. Thus, we have to insert Eq. \ref{identity} into Eq. \ref{flux}. We have also used the fact that the $S-$matrix involving $\ket{p_{2i}}$ will be proportional to $\sqrt{\kappa_{2i}}$, by replacing $\sqrt{\kappa_{2e}}$ in Eq. \ref{Sp3}.  Exchanging subscripts 1 and 2 in Eq. \ref{flux} yields the result of $\bra{\psi_{\textrm{in}}}a_{\textrm{out},2}^\dagger(t)a_{\textrm{out},2}(t)\ket{\psi_{\textrm{in}}}$. The intra-cavity photon-pair generation rate is inferred from Eq. \ref{flux} to be
\be
R=|\alpha|^2\frac{g^2(\kappa_1+\kappa_2)\kappa_{3e}}{|(k-\lambda_1)(k-\lambda_2)|^2}.
\ee
 
Finally, we have
\begin{equation}\begin{aligned}
g^{(2)}(\tau)=&\frac{|(k-\lambda_1)(k-\lambda_2)|^2}{2|\alpha|^2g^2(\kappa_1+\kappa_2)\kappa_{3e}}\frac{2\kappa_1\kappa_2}{\kappa_1+\kappa_2}\times\begin{cases}
e^{-\kappa_2\tau}, \tau>0,\\
e^{\kappa_1\tau}, \tau<0,
\end{cases}
\end{aligned}\end{equation}
and in the limit $g\ll \kappa_{1,2,3}$,
\begin{equation}\begin{aligned}\label{downg2result}
g^{(2)}(\tau)\approx&\frac{\left((k-\omega_1-\omega_2)^2+(\frac{\kappa_1+\kappa_2}{2})^2\right)\left((k-\omega_3)^2+(\frac{\kappa_3}{2})^2\right)}{2|\alpha|^2g^2(\kappa_1+\kappa_2)\kappa_{3e}}\\
&\times\frac{2\kappa_1\kappa_2}{\kappa_1+\kappa_2}\times\begin{cases}
e^{-\kappa_2\tau}, \tau>0,\\
e^{\kappa_1\tau}, \tau<0.
\end{cases}
\end{aligned}\end{equation}
Interestingly, Eq. \ref{downg2result} is identical to the result of spontaneous parametric down-conversion derived via a semi-classical approach by assuming non-depleted classical pumps and using Gaussian moment factoring theorem \cite{clausen2014source,guo2017parametric}.

\subsection{Photon blockade}\label{Subsection:IVB}

Next, we consider transport of $a_1$ photons via a cavity with degenerate $a_1$ and $a_2$ modes. For input of $a_1$ photons in this case, there will be probability of up-conversion to output $a_3$ photons; however, we focus on the process without such up-conversion, which could be separated apart from the up-converted $a_3$ photons because $a_1$ and $a_3$ photons are disparate in frequency. Assuming monochromatic input $a_1$ photons in a weak coherent state given by Eq. \ref{weakalpha}, we compute the second-order self-correlation function of the transported $a_1$ photons:
\begin{equation}
\begin{aligned}\label{g2smdef}
&g^{(2)}(\tau)\\=&\frac{\bra{\psi_{\textrm{in}}}a_{\textrm{out},1}^\dagger(t)a_{\textrm{out},1}^\dagger(t+\tau)a_{\textrm{out},1}(t+\tau)a_{\textrm{out},1}(t)\ket{\psi_{\textrm{in}}}}{\bra{\psi_{\textrm{in}}}a_{\textrm{out},1}^\dagger(t)a_{\textrm{out},1}(t)\ket{\psi_{\textrm{in}}}\bra{\psi_{\textrm{in}}}a_{\textrm{out},1}^\dagger(t)a_{\textrm{out},1}(t)\ket{\psi_{\textrm{in}}}}.
\end{aligned}
\end{equation}
By keeping the leading order terms in $\ket{\psi_{\textrm{in}}}$, we have
\begin{equation}\begin{aligned}
&\bra{\psi_{\textrm{in}}}a_{\textrm{out},1}^\dagger(t)a_{\textrm{out},1}(t)\ket{\psi_{\textrm{in}}}\\
=&|\alpha|^2\bra{1_t} a_{\textrm{out},1}^\dagger(t)a_{\textrm{out},1}(t)\ket{1_t}\\
=&|\alpha|^2\bra{1_t} a_{\textrm{out},1}^\dagger(t)\ket{0}\bra{0}a_{\textrm{out},1}(t)\ket{1_t}\\
=&|\alpha|^2\int dp e^{ipt}\bra{1_k} a_{\textrm{out},1}^\dagger(p)\ket{0} \int d\tilde p e^{-i\tilde pt}\bra{0}a_{\textrm{out},1}(\tilde p)\ket{1_k}\\
=&|\alpha|^2\int dp e^{ipt}S_{p;k}^* \int d\tilde p e^{-i\tilde pt}S_{\tilde p;k} \\
=&|\alpha|^2|t_k|^2
\end{aligned}\end{equation}
and
\begin{equation}\begin{aligned}\label{g2num}
&\bra{\psi_{\textrm{in}}}a_{\textrm{out},1}^\dagger(t)a_{\textrm{out},1}^\dagger(t+\tau)a_{\textrm{out},1}(t+\tau)a_{\textrm{out},1}(t)\ket{\psi_{\textrm{in}}}\\
=&\frac{|\alpha|^4}{2}\bra{1_t1_t}a_{\textrm{out},1}^\dagger(t)a_{\textrm{out},1}^\dagger(t+\tau)a_{\textrm{out},1}(t+\tau)a_{\textrm{out},1}(t)\ket{1_t1_t}\\
=&\frac{|\alpha|^4}{2}\int dp_1dp_2 \bra{1_k1_k}a_{\textrm{out},1}^\dagger(p_1)a_{\textrm{out},1}^\dagger(p_2)\ket{0}  e^{ip_1t+ip_2(t+\tau)}\\
&\quad\times\int d\tilde p_1d\tilde p_2\bra{0}a_{\textrm{out},1}(\tilde p_2)a_{\textrm{out},1}(\tilde p_1)\ket{1_k1_k}e^{-i\tilde p_1t-i\tilde p_2(t+\tau)}\\
=& \frac{|\alpha|^4}{4}\int dp_1dp_2 S_{p_1p_2;kk}^* e^{ip_1t+ip_2(t+\tau)}\\
&\quad \times \int d\tilde p_1d\tilde p_2 S_{\tilde p_1\tilde p_2;kk}e^{-i\tilde p_1t-i\tilde p_2(t+\tau)} \\
=&|\alpha|^4|t_k^2+T(k,\tau)|^2,
\end{aligned}\end{equation}
where the $S-$matrix is given by Eq. \ref{S22d} and
\begin{equation}\begin{aligned}\label{Tktaud}
T(k,\tau)=-\frac{2g^2\kappa_{1e}^2}{(2k-\lambda'_1)(2k-\lambda'_2)(k-\alpha_1)^2}e^{-i|\tau|(\alpha_1-k)}.
\end{aligned}\end{equation}
Note the additional factor of 1/2 in the third equality of Eq. \ref{g2num} comes from the definition of Fock state $\ket{1_k1_k}=\ket{2_k}=\frac{1}{\sqrt{2}}a_{\textrm{in}}^{\dagger 2}(k)\ket{0}$. 
Finally, we have
\begin{equation}\begin{aligned}\label{g2result}
g^{(2)}(\tau)=\frac{|t_k^2+T(k,\tau)|^2}{|t_k^2|^2}.
\end{aligned}\end{equation}

\begin{figure}[!tb]
\includegraphics[width=0.5\textwidth]{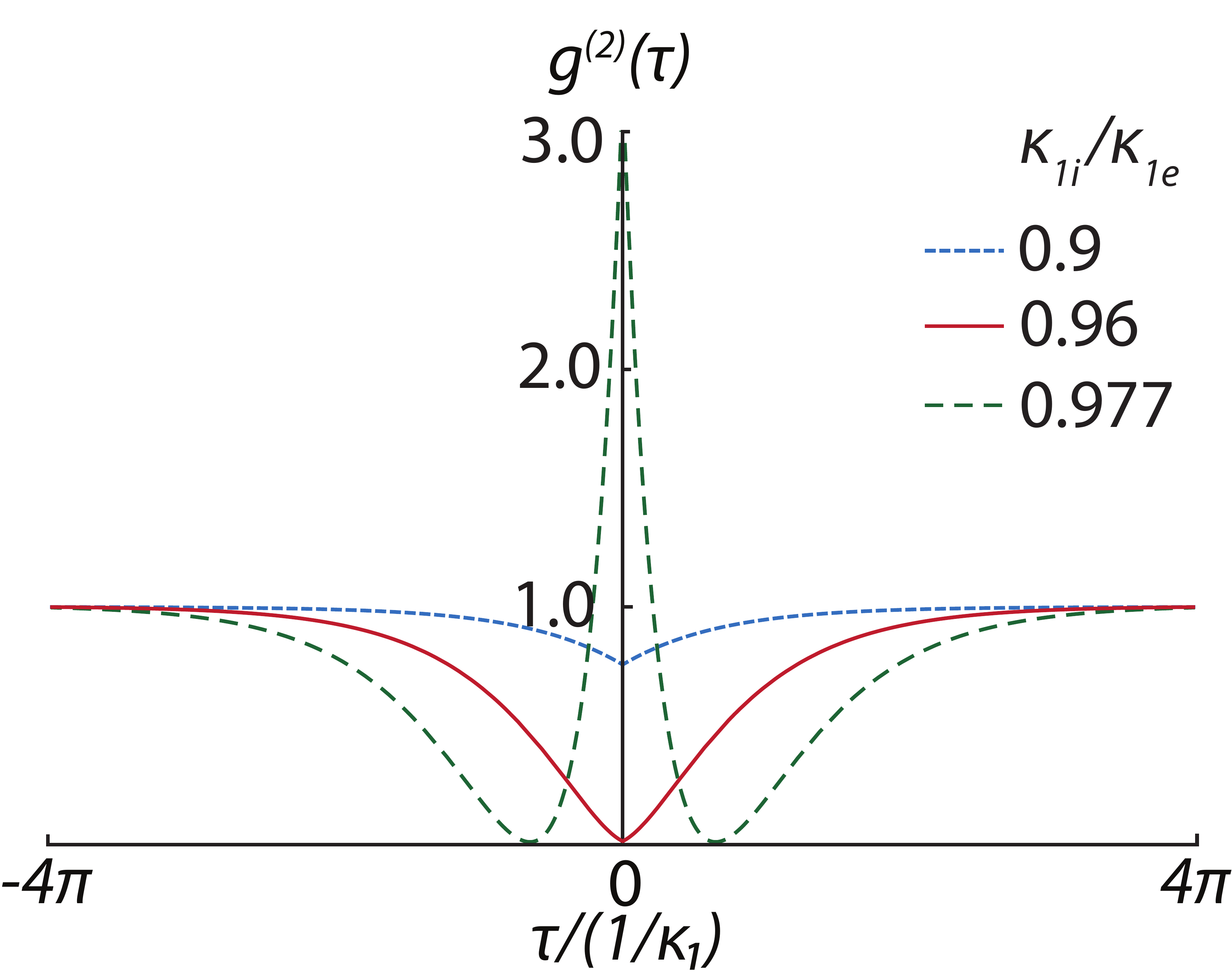}
\caption{$g^{(2)}(\tau)$ for various $\kappa_{1e}/\kappa_{1i}$, given $k=\omega_1$, $g/\kappa_{1i}=0.02$, $\kappa_{3i}=2\kappa_{1i}$, $\kappa_{3e}=0.1\kappa_{1i}$.}\label{fig:g2}
\end{figure}

This result is remarkable in that even when $|T(k,\tau)|$ is small, which is the case for practical $\chi^{(2)}$ nonlinear optical cavities with $g/\kappa\ll 1$, highly nonclassical correlations of transported photons, i.e., $g^{(2)}(\tau)\neq 1$, could be achieved by making $|t_k^2|$ close to $|T(k,\tau)|$. This is due to the quantum interference between the interaction-free and interaction-mediated amplitudes of the two-photon transport, as indicated by the $S-$matrix (Eq. \ref{S22}), especially when the two amplitudes are compatible. To illustrate this more explicitly, for the one-port, phase-matched cavity ($2\omega_1=\omega_3$) and on-resonance input photons ($k=\omega_1$), 
\be
t_{\omega_1}^2=(\frac{\kappa_{1i}-\kappa_{1e}}{\kappa_{1i}+\kappa_{1e}})^2
\ee
and 
\be
T(\omega_1,0)=-\frac{8g^2\kappa_{1e}^2}{\kappa_1^2(\kappa_1\kappa_3/2+2g^2)}.
\ee
Thus, by making $\kappa_{1i}$ and $\kappa_{1e}$ such that $t_{\omega_1}^2\approx -T(\omega_1,0)$, one obtains $g^{(2)}(0)\approx 0$, which indicates the photon blockade.  Remarkably, this can be achieved even for $g/\kappa_{1,3}\ll 1$, if $\kappa_{1i}\approx\kappa_{1e}$, i.e., close to the critical coupling between the cavity and waveguide. We note the minus sign of $T(\omega_1,0)$ is important in cancellation with $t_{\omega_1}^2$; intuitively, this is because during the interaction-mediated process the state $\ket{20}$ undergoes a Rabi flip with the virtual state $\ket{01}$ and back again, which yields a Berry phase of $\pi$. 

Moreover, the statistical properties of the output $a_1$ photons can be altered, e.g., from anti-bunching to bunching ($g^{(2)}(0)>g^{(2)}(\tau)$), by controlling $t_k^2$ via $\kappa_{1e}$ or $\kappa_{1i}$. This is shown in Fig. \ref{fig:g2}. For these plots, we have used experimentally achievable device parameters of a state-of-the-art $\chi^{(2)}$ nonlinear photonic platform with $g/\kappa_{1i}$ in the range of a few percent \cite{zhao2021ingap}. The correlation functions of finite wavepackets can also be calculated using the $S$-matrix, taking into account of the spectral distribution of the wavepacket. The result is included in  Appendix~\ref{Appendix:B}, which shows the averaging effect on the nonclassical correlation due to the finite spectral width of the wavepacket. However, this is not surprising because the cavity-based mechanism is expected to work only for wavepackets with spectral width much smaller than the cavity linewidth.

\subsection{Nonlinear phase shift}\label{Subsection:IVC}

\begin{figure}[!tb]
\includegraphics[width=0.5\textwidth]{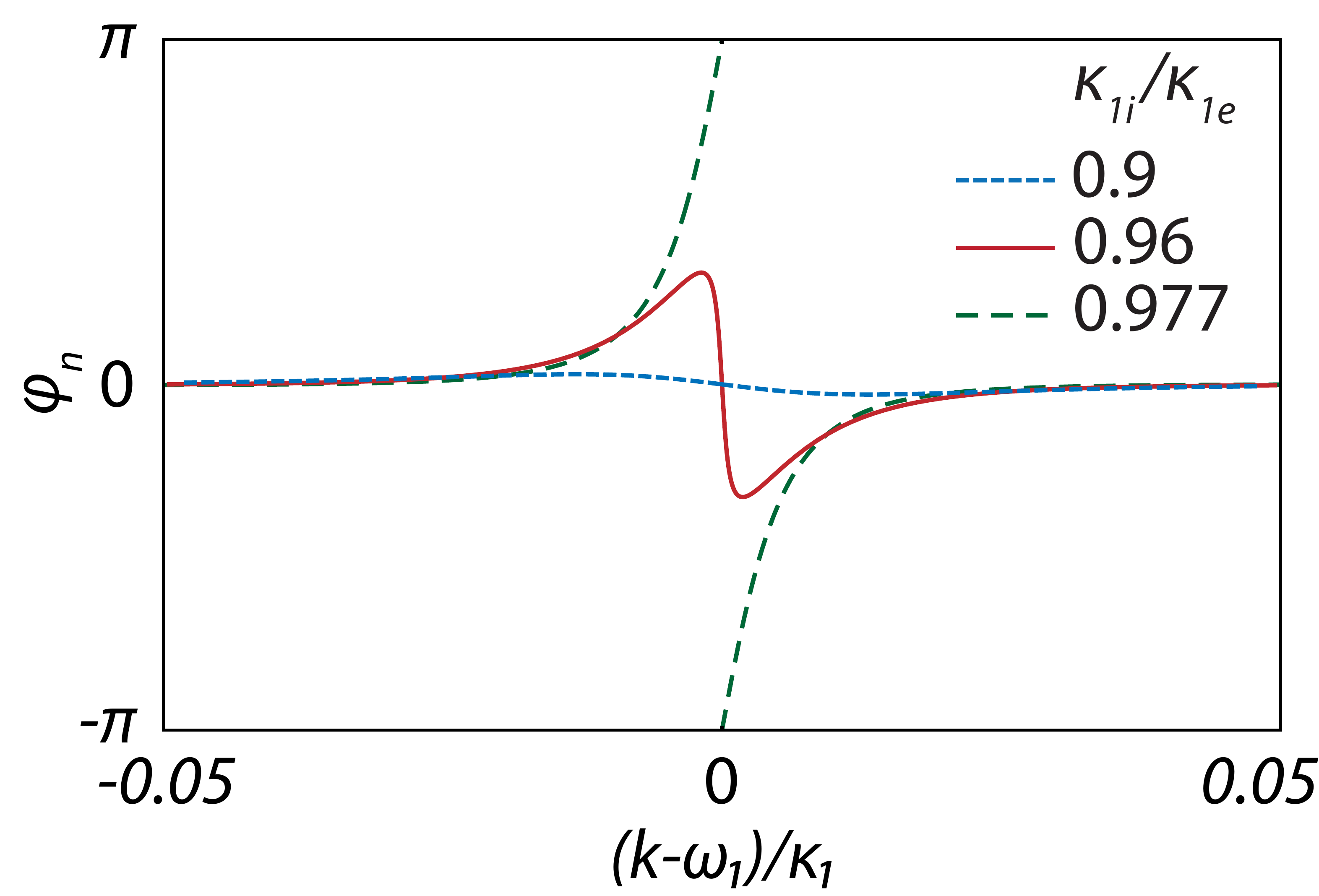}
\caption{Nonlinear phase shift for various $\kappa_{1e}/\kappa_{1i}$, given $g/\kappa_{1i}=0.02$, $\kappa_{3i}=2\kappa_{1i}$, $\kappa_{3e}=0.1\kappa_{1i}$. }\label{cphase}
\end{figure}

The photon-photon interaction in the $\chi^{(2)}$ cavity also induces a nonlinear phase shift of the transported photons. To compute the nonlinear phase shift, we point out the numerator of the second-order correlation function (Eq. \ref{g2smdef}) actually gives the norm of the time-domain wavefunction of the transported two photons (see Appendix \ref{Appendix:A}). Thus, the wavefunction of the transported two photons can be written as
\begin{equation}
\bra{t,t+\tau}\ket{\psi_{\textrm{out}}}=e^{-ik(2t+\tau)}(t_k^2+T(k,\tau)).
\end{equation}
The wavefunction of two photons without the interaction is
\begin{equation}
\bra{t,t+\tau}\ket{\psi_{\textrm{out}}^{(0)}}=e^{-ik(2t+\tau)}t_k^2.
\end{equation}
The nonlinear phase shift due to the photon-photon interaction thus is 
\begin{equation}\begin{aligned}
\varphi_{\textrm{n}}&\equiv \textrm{Arg}\left[\bra{t,t+\tau}\ket{\psi_{\textrm{out}}}\right]-\textrm{Arg}\left[\bra{t,t+\tau}\ket{\psi_{\textrm{out}}^{(0)}}\right]\\
&= \textrm{Arg}\left[1+\frac{T(k,\tau)}{t_k^2}\right].
\end{aligned}\end{equation}
By making $|t_k^2|\approx|T(k,\tau)|$, large nonlinear phase shift can be achieved, which is possible even for the weak coupling regime $g\ll \kappa_1$. Such nonlinear phase shift is a key enabler for quantum nondemolition measurement of photons \cite{xia2016cavity} and conditional quantum logic gates \cite{turchette1995measurement,langford2011efficient}.

Fig. \ref{cphase} shows the calculated nonlinear phase shift (for $\tau=0$) using experimentally achievable parameters of a state-of-the-art $\chi^{(2)}$ nonlinear photonic platform \cite{zhao2021ingap}. It is seen that large and even $\pi-$nonlinear phase shift can be achieved in practical devices. Actually, $\pi-$nonlinear phase shift is always achieved at $k=\omega_1$ as long as $t_{\omega_1}^2<-T(\omega_1,0)$.

\subsection{Two-mode nonclassical correlation}
In Sections \ref{Subsection:IVB} and \ref{Subsection:IVC}, we considered a cavity with degenerate $a_1$ and $a_2$ modes. Here, we consider a cavity with non-degenerate $a_1$ and $a_2$ modes which satisfy the frequency-matching condition $\omega_1+\omega_2\approx \omega_3$ and show nonclassical two-mode correlations could be generated. We assume now the input state to be the product of two weak coherent states in modes $a_1$ and $a_2$,
\begin{equation}
\begin{aligned}
\ket{\psi_{\textrm{in}}}=&\ket{\alpha}_1\ket{\beta}_2\\=&\left(\ket{0}+\alpha\ket{1_{t}}_1+\frac{\alpha^2}{\sqrt{2}}\ket{1_{t}1_{t}}_1+O(\alpha^3)\right)\\&\times\left(\ket{0}+\beta\ket{1_{t}}_2+\frac{\beta^2}{\sqrt{2}}\ket{1_{t}1_{t}}_2+O(\beta^3)\right).
\end{aligned}
\end{equation}
The two-mode and single-mode correlations are, respectively,
\begin{equation}\begin{aligned}
&G_{12}(\tau)\\=&\bra{\psi_{\textrm{in}}}a_{\textrm{out},1}^\dagger(t)a_{\textrm{out},2}^\dagger(t+\tau)a_{\textrm{out},2}(t+\tau)a_{\textrm{out},1}(t)\ket{\psi_{\textrm{in}}}\\
=&|\alpha\beta|^2\int dp_1dp_2 \bra{1_{k_1}1_{k_2}}a_{\textrm{out},1}^\dagger(p_1)a_{\textrm{out},2}^\dagger(p_2)\ket{0}  e^{ip_1t+ip_2(t+\tau)}\\
&\quad\times\int d\tilde p_1d\tilde p_2\bra{0}a_{\textrm{out},2}(\tilde p_2)a_{\textrm{out},1}(\tilde p_1)\ket{1_{k_1}1_{k_2}}e^{-i\tilde p_1t-i\tilde p_2(t+\tau)}\\
=& |\alpha\beta|^2\int dp_1dp_2 S_{p_1p_2;k_1k_2}^* e^{ip_1t+ip_2(t+\tau)}\\
&\quad \times \int d\tilde p_1d\tilde p_2 S_{\tilde p_1\tilde p_2;k_1k_2}e^{-i\tilde p_1t-i\tilde p_2(t+\tau)} \\
=&|\alpha\beta|^2|t_{k_1}t_{k_2}+\widetilde{T}(k_1,k_2,\tau)|^2,
\end{aligned}\end{equation}
where the $S-$matrix is given by Eq. \ref{S22} and 
\begin{equation}\begin{aligned}\label{2modeT}
&\widetilde{T}(k_1,k_2,\tau)\\
&=
\begin{cases}
-\frac{g^2\kappa_{1e}\kappa_{2e}}{(k_1+k_2-\lambda_1)(k_1+k_2-\lambda_2)(k_1-\alpha_1)(k_2-\alpha_2)}e^{-i\tau(\alpha_2-k_2)}, \tau>0,\\ 
-\frac{g^2\kappa_{1e}\kappa_{2e}}{(k_1+k_2-\lambda_1)(k_1+k_2-\lambda_2)(k_1-\alpha_1)(k_2-\alpha_2)}e^{i\tau(\alpha_1-k_1)}, \tau<0,
\end{cases}
\end{aligned}\end{equation}
and 
\begin{equation}\begin{aligned}\label{G11}
&G_{11}(\tau)\\=&\bra{\psi_{\textrm{in}}}a_{\textrm{out},1}^\dagger(t)a_{\textrm{out},1}^\dagger(t+\tau)a_{\textrm{out},1}(t+\tau)a_{\textrm{out},1}(t)\ket{\psi_{\textrm{in}}}\\
=&\frac{|\alpha|^4}{2}\int dp_1dp_2 \bra{1_{k_1}1_{k_1}}a_{\textrm{out},1}^\dagger(p_1)a_{\textrm{out},1}^\dagger(p_2)\ket{0}  e^{ip_1t+ip_2(t+\tau)}\\
&\quad\times\int d\tilde p_1d\tilde p_2\bra{0}a_{\textrm{out},1}(\tilde p_1)a_{\textrm{out},1}(\tilde p_2)\ket{1_{k_1}1_{k_1}}e^{-i\tilde p_2t-i\tilde p_1(t+\tau)}\\
=& \frac{|\alpha|^4}{4}\int dp_1dp_2 2t_{k_1}^*t_{k_1}^*\delta(p_1-k_1)\delta(p_2-k_1) e^{ip_1t+ip_2(t+\tau)}\\
&\quad \times \int d\tilde p_1d\tilde p_2 2t_{k_1}t_{k_1}\delta(p_1-k_1)\delta(p_2-k_1)e^{-i\tilde p_2t-i\tilde p_1(t+\tau)} \\
=&|\alpha|^4|t_{k_1}|^4,
\end{aligned}\end{equation}
and similarly,
\be
G_{22}(\tau)=|\beta|^4|t_{k_2}|^4.
\ee
In Eq. \ref{G11}, we have used the fact that the transport of two $a_1$ photons via a non-degenerate cavity is interaction-free because $2\omega_1\neq\omega_3$.

To measure the nonclassicality of the two-mode correlation, we define
\be\label{zeta}
\zeta(\tau)\equiv\frac{G_{12}(\tau)}{\sqrt{G_{11}(\tau)G_{22}(\tau)}}=|1+\frac{\widetilde{T}(k_1,k_2,\tau)}{t_{k_1}t_{k_2}}|^2.
\ee
$\zeta(\tau)>1$ leads to violation of the Cauchy-Schwartz inequality and indicates nonclassical two-mode correlations \cite{mandel1995optical}. 
Given the similarity of Eq. \ref{zeta} and Eq. \ref{g2result}, it is obvious that $\zeta(\tau)>1$ can be achieved when $|t_{k_1}t_{k_2}|$ is sufficiently smaller than $|\widetilde{T}(k_1,k_2,\tau)|$.

\section{Conclusion}
In summary, we have theoretically studied few-photon transport via a waveguide-coupled, multimode $\chi^{(2)}$ optical cavity. We used both non-perturbative method and a new perturbation method based on Feynman diagrams to compute the $S-$matrix associated with the few-photon transport. The Feynman diagram approach provides physical insight into the transport process and is mathematically convenient. The $S-$matrix shows that the two-photon transport involves quantum interference between linear transmission and interaction-mediated transport. This effect leads to rather unexpected result that strong quantum correlations could be realized in weak nonlinear systems by matching the linear transmission coefficient with the interaction-mediated amplitude. We numerically showed several pronounced quantum optical effects, including photon blockade and $\pi-$conditional phase shift, could be achieved in state-of-the-art nonlinear quantum photonic platforms, which might have a significant impact on using these systems for quantum information science applications.

\appendix
\section{The wavefunction of transported photons}\label{Appendix:A}
Here we prove the second-order correlation actually gives the norm of the output two-photon wavefunction.
\begin{widetext}
\begin{equation}\begin{aligned}
&\bra{\psi_{\textrm{in}}}a_{\textrm{out}}^\dagger(t)a_{\textrm{out}}^\dagger(t+\tau)a_{\textrm{out}}(t+\tau)a_{\textrm{out}}(t)\ket{\psi_{\textrm{in}}}\\
=&\frac{1}{(2\pi)^2}\int dp_1dp_2dp_3dp_4e^{i(p_1t+p_2(t+\tau)-p_3(t+\tau)-p_4t)}\bra{\psi_{\textrm{in}}}\hat S^\dagger\hat Sa_{\textrm{out}}^\dagger(p_1)\hat S^\dagger\hat Sa_{\textrm{out}}^\dagger(p_2)\hat S^\dagger\hat Sa_{\textrm{out}}(p_3)\hat S^\dagger\hat Sa_{\textrm{out}}(p_4)\hat S^\dagger\hat S\ket{\psi_{\textrm{in}}}\\
=&\frac{1}{(2\pi)^2}\int dp_1dp_2dp_3dp_4e^{i(p_1t+p_2(t+\tau)-p_3(t+\tau)-p_4t)}\bra{\psi_{\textrm{out}}}a_{\textrm{in}}^\dagger(p_1)a_{\textrm{in}}^\dagger(p_2)a_{\textrm{in}}(p_3)a_{\textrm{in}}(p_4)\ket{\psi_{\textrm{out}}}\\
=&\frac{1}{(2\pi)^2}\int dp_1dp_2dp_3dp_4e^{i(p_1t+p_2(t+\tau)-p_3(t+\tau)-p_4t)}\bra{\psi_{\textrm{out}}}a_{\textrm{in}}^\dagger(p_1)a_{\textrm{in}}^\dagger(p_2)\ket{0}\bra{0}a_{\textrm{in}}(p_3)a_{\textrm{in}}(p_4)\ket{\psi_{\textrm{out}}}\\
=&\frac{1}{(2\pi)^2}\int dp_1dp_2e^{i(p_1t+p_2(t+\tau))}\bra{p_1,p_2}\ket{\psi_{\textrm{out}}}^*\int dp_3dp_4e^{-i(p_3(t+\tau)+p_4t)}\bra{p_3,p_4}\ket{\psi_{\textrm{out}}}\\
=&|\bra{t,t+\tau}\ket{\psi_{\textrm{out}}}|^2
\end{aligned}\end{equation}
\end{widetext}
where $\ket{t,t+\tau}$ represents the state of two photons at time $t$ and $t+\tau$, respectively, and thus the end result gives the time-domain wavefunction of the output two photons. We have used $\hat S$ to represent the operator corresponding to the $S$-matrix defined in the momentum space, which satisfies the unitary condition $\hat S^\dagger\hat S=\hat I$. We have also assumed the input/output states contain up to two photons, which is consistent with Eq. \ref{weakalpha}. 

\section{Correlation functions for photon wavepackets}\label{Appendix:B}
\begin{figure}[!tb]
\includegraphics[width=0.46\textwidth]{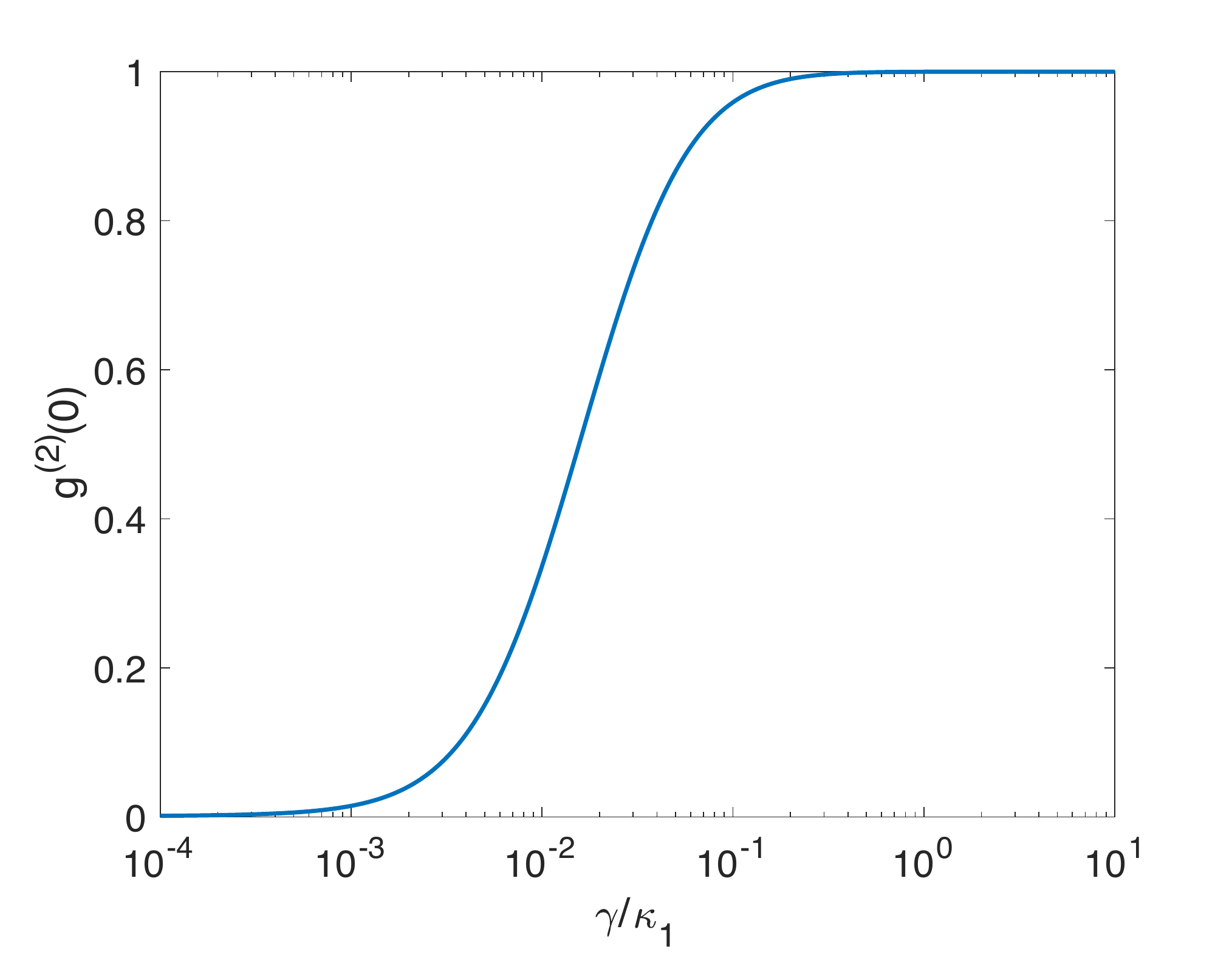}
\caption{ $g^{(2)}(0)$ as a function of wavepacket width $\gamma$.  Other parameters used in the calculation are  $k_0=\omega_1$, $g/\kappa_{1i}=0.02$,  $\kappa_{1i}/\kappa_{1e}=1.04$, $\kappa_{3i}=2\kappa_{1i}$, $\kappa_{3e}=0.1\kappa_{1i}$. }\label{wavepacket}
\end{figure}

We consider the input state being a photon wavepacket, i.e., $\ket{\psi_{\text{in}}}=\exp(\alpha a^\dagger-\alpha^* a)\ket{0}$, where $a=\int dk f(k) a_k$ and $f(k)$ is the spectral distribution centered around $k_0$ and satisfying $\int |f(k)|^2 dk=1$. The first- and second-order correlation functions for the photon wavepacket are given by 
\begin{widetext}
\begin{equation}\begin{aligned}
&\bra{\psi_{\textrm{in}}}a_{\textrm{out},1}^\dagger(t)a_{\textrm{out},1}(t)\ket{\psi_{\textrm{in}}}\\
=&|\alpha|^2\int dk_1 d\tilde k_1 f(k_1)f(\tilde k_1)\bra{1_{k_1}} a_{\textrm{out},1}^\dagger(t)a_{\textrm{out},1}(t)\ket{1_{\tilde  k_1}}\\
=&|\alpha|^2\int dk_1 d\tilde k_1 f(k_1)f(\tilde k_1)\bra{1_{k_1}} a_{\textrm{out},1}^\dagger(p_1)a_{\textrm{out},1}(\tilde p_1)\ket{1_{\tilde k_1}}e^{i(p_1-\tilde p_1)t}\\
=&|\alpha|^2\int dk_1 dp_1 f(k_1)e^{ip_1t}S^*_{p_1k_1}\int d\tilde k_1 d\tilde p_1 f(\tilde k_1)e^{-i\tilde p_1t}S_{\tilde p_1\tilde k_1}\\
=&|\alpha|^2\int dk_1  f(k_1)e^{ik_1t}t_{k_1}^* \int d\tilde k_1  f(\tilde k_1)e^{-i\tilde k_1t}t_{\tilde k_1}\\
=&|\alpha|^2|\widetilde t(k_0,t)|^2,
\end{aligned}\end{equation}
where $\widetilde t(k_0,t)=\int d k_1  f( k_1)e^{-i k_1t}t_{ k_1}$, and
\begin{equation}\begin{aligned}
&\bra{\psi_{\textrm{in}}}a_{\textrm{out},1}^\dagger(t)a_{\textrm{out},1}^\dagger(t+\tau)a_{\textrm{out},1}(t+\tau)a_{\textrm{out},1}(t)\ket{\psi_{\textrm{in}}}\\
=&\frac{|\alpha|^4}{2}\int dp_1dp_2dk_1 dk_2 \bra{1_{k_1}1_{k_2}}a_{\textrm{out},1}^\dagger(p_1)a_{\textrm{out},1}^\dagger(p_2)\ket{0}  e^{ip_1t+ip_2(t+\tau)}f(k_1)f(k_2)\\
&\quad\times\int d\tilde p_1d\tilde p_2 d\tilde k_1d\tilde k_2\bra{0}a_{\textrm{out},1}(\tilde p_2)a_{\textrm{out},1}(\tilde p_1)\ket{1_{\tilde k_1}1_{\tilde k_2}}e^{-i\tilde p_1t-i\tilde p_2(t+\tau)}f(\tilde k_1)f(\tilde k_2)\\
=& \frac{|\alpha|^4}{4}\int dp_1dp_2 S_{p_1p_2;k_1k_1}^* e^{ip_1t+ip_2(t+\tau)}f(k_1)f(k_2)\\
&\quad \times \int d\tilde p_1d\tilde p_2 S_{\tilde p_1\tilde p_2;\tilde  k_1 \tilde k_2}e^{-i\tilde p_1t-i\tilde p_2(t+\tau)}f(\tilde k_1)f(\tilde k_2) \\
=&|\alpha|^4|\widetilde t(k_0,t) \widetilde t(k_0,t+\tau)+\widetilde T(k_0,t,\tau)|^2,
\end{aligned}\end{equation}
where
\begin{equation}\begin{aligned}
\widetilde T(k_0,t,\tau)=\frac{1}{2}\int dk_1dk_2dp_1\kappa_{1e}^2M(p_1,k_1+k_2-p_,k_1,k_2)e^{-ip_1t-i(k_1+k_2-p_1)(t+\tau)}f(k_1)f(k_2).
\end{aligned}\end{equation}

To illustrate the effect of finite spectral width of wavepackets, we consider a spectral distribution of $f(k)=\sqrt{\frac{2}{\pi\gamma}}\frac{\gamma^2}{(k-k_0)^2+\gamma^2}$ which leads to
\begin{equation}\begin{aligned}
&\widetilde t(k_0,t)=
\begin{cases}
\sqrt{2\pi \gamma}[e^{-i(k0-i\gamma)t}(1-\frac{i\kappa_{1e}}{k_0-i\gamma-\alpha_1})-e^{-i\alpha_1 t}\frac{2\gamma\kappa_{1e}}{(\alpha_1-k_0)^2+\gamma^2}],\quad t>0,\\ 
\sqrt{2\pi \gamma}e^{-i(k0+i\gamma)t}(1-\frac{i\kappa_{1e}}{k_0+i\gamma-\alpha_1}),\quad t<0
\end{cases}
\end{aligned}\end{equation}
and
\begin{equation}\begin{aligned}
&\widetilde T(k_0,t=0,\tau)= -\frac{4\pi g^2 \kappa_{1e}^2\gamma}{(k_0+i\gamma-\alpha_1)^2(2k_0+2i\gamma-\lambda_1')(2k_0+2i\gamma-\lambda_2')}\times
\begin{cases}
e^{-i\alpha_1\tau},\quad \tau>0,\\ 
e^{-2i(k_0+i\gamma)\tau+i\alpha_1\tau},\quad \tau<0.
\end{cases}
\end{aligned}\end{equation}
\end{widetext}
We assume the wavepacket to be sufficiently long and thus ignore the temporal wavepacket shape. The normalized second-order correlation function thus is given by, setting $t=0$,
\begin{equation}\begin{aligned}\label{g2result}
g^{(2)}(\tau)=\frac{|\widetilde t(k_0,0) \widetilde t(k_0,\tau)+\widetilde T(k_0,0,\tau)|^2}{|\widetilde t(k_0,0)|^2|\widetilde t(k_0,\tau)|^2}.
\end{aligned}\end{equation}
We plot $g^{(2)}(0)$ as a function of $\gamma$ in Fig.~\ref{wavepacket} for a parameter set which yields $g^{(2)}(0)=0$ for $\gamma\rightarrow 0$, i.e., the continuous-wave limit. The averaging effect is observed due to the spectral span of the photon wavepacket, leading to $g^{(2)}(0)$ increasing from 0 to 1. However, $g^{(2)}(0)$ remains sufficiently small for $\gamma\leq 10^{-2}\kappa_1$.

%

\vspace{2mm}
\noindent\textbf{Acknowledgements}\\ 
This work is supported by US National Science Foundation under Grant No. DMS 18-39177.

\end{document}